\documentclass[pdftex,10pt,a4paper]{article}

\usepackage{booktabs} 
\usepackage{multirow}
\usepackage{soul} 
\usepackage{microtype}
\usepackage{upgreek}
\usepackage{graphicx}

\begin{document}

\thispagestyle{plain}
\begin{center}
    {\Large
    \textbf{The Connection between Biaxial Orientation and Shear Thinning for Quasi-Ideal Rods}}
\end{center}
    \vspace{0.4cm}
    \textbf{Christian Lang $^{1,}$*, Joachim Kohlbrecher $^{2}$, Lionel Porcar $^{3}$ and Minne Paul Lettinga $^{1,\dagger}$ }

\vspace{0.4cm}
\noindent$^{1}$ ICS-3, Institut f\"{u}r Weiche Materie, Forschungszentrum J\"{u}lich, D-52425 J\"{u}lich, Germany; p.lettinga@fz-juelich.de \\
\noindent $^{2}$ Laboratory of Neutron Scattering and Imaging, Paul Scherrer Institute, 5232 Villigen, Switzerland; joachim.kohlbrecher@psi.ch\\ 
\noindent$^{3}$ Institut Laue-Langevin, CS 20156, 38042 Grenoble Cedex 9, France; porcar@ill.fr\\
\noindent*  Correspondence: c.lang@fz-juelich.de; Tel.: +49-246-161-2149\\
\noindent$^\dagger$ Also at: Laboratory for Soft Matter and Biophysics, KU Leuven, Celestijnenlaan 200D, B-3001 Leuven, Belgium

 \vspace{0.9cm}
\noindent    \textbf{Abstract}
The complete orientational ordering tensor of quasi-ideal colloidal rods is obtained as a function of shear rate by performing rheo-SANS (rheology with small angle neutron scattering) measurements on isotropic fd-virus suspensions in the two relevant scattering planes, the flow-gradient (1-2) and the flow-vorticity (1-3) plane. Microscopic ordering can be identified as the origin of the observed shear thinning. 
A qualitative description of the rheological response by Smoluchowski, as well as Doi--Edwards--Kuzuu theory is possible, as we obtain a master curve for different concentrations, scaling the shear rate with the apparent collective rotational diffusion coefficient.
However, the observation suggests that the interdependence of ordering and shear thinning at small shear rates is stronger than predicted. The extracted zero-shear viscosity matches the concentration dependence of the self-diffusion of rods in semi-dilute solutions, while the director tilts close towards the flow direction already at very low shear rates. In contrast, we observe a smaller dependence on the shear rate in the overall ordering at high shear rates, as well as an ever-increasing biaxiality.

\vspace{0.4cm}

\section{Introduction}
\label{1}

The connection between the macroscopic rheological properties of semi-flexible or stiff polymer dispersions and the underlying microstructural changes is of high fundamental importance and also plays a huge role in many industrial processes. As a starting point for understanding semi-flexible systems, one can investigate a quasi-ideal stiff polymer suspension and define its rheological behavior, as well as its structure in flow. This benchmark problem could help to disentangle the manifold contributions to the complex flow-response of semi-flexible systems and make further progress in their theoretical description.
Shear thinning is one of the main rheological signatures of a complex structural response of a material to shear flow. Generally, this type of non-linear response is probed by steady shear experiments where an increasing shear rate is applied and a stress is measured or vice versa. For a fundamental understanding of shear thinning in terms of the microstructural changes of sheared material, a technique is needed that probes the sample on the relevant length-scales. Moreover, model systems are needed that display very reproducible flow behavior and which can ideally also be modeled with the appropriate microscopic theory. The goal of this paper is exactly this, namely to study the connection between the microstructural and mechanical response of the model rod-like fd-virus viruses, combining stationary shear rheology with small angle neutron scattering (SANS), so-called rheo-SANS.

Fd-virus has been extensively used for fundamental studies during the last few decades \cite{BaRu+1994,GoSc2002}, as their high aspect-ratio and mono-dispersity makes them a quasi-ideal system for investigating the equilibrium \cite{DoFr2006,Grel2014} and the out-of-equilibrium behavior \cite{RiHo+2008,LeDh2001, LeDh2004,TaFr1995,GrKr+1992} of rod-like colloids. 
It has been used, for~example, to test the theoretically-predicted dynamical transitions in sheared nematics~\cite{Hess1976,HeKr2004,KlHe2010}. For those systems, indeed, a quantitative comparison with simulations \cite{TaDe2006} and theory~\cite{LeDo+2005} was found.
Though the isotropic phase in shear flow does not display as rich dynamic transitions as the nematic phase, the understanding of its flow behavior is complicated. The reason is that shear flow competes with the reorientational motion of the rods, which has a collective contribution and a contribution due to the mobility of a single particle. Scaling the shear rate, that is defining the right Peclet number, is not straightforward, as collective diffusion depends on the self-diffusion~\cite{TaDe2006}. Smoluchowski-based theory \cite{DhBr2003,OlLu1999} predicts that shear flow will strongly affect the orientation when approaching the isotropic-nematic transition, as here, the collective diffusion goes to zero, which is confirmed by experiments on fd-virus \cite{LeDh2001}.
However, shear cannot induce a real phase transition out of an initially isotropic phase, as was shown by a combination of simulations and experiments on fd-virus, because of the markedly different response of the nematic and isotropic phase \cite{RiHo+2008}.

The disagreement between the measured shear dependent viscosity by Graf et al. \cite{GrKr+1992} and Smoluchowski equation-based theory \cite{DhBr2003} is apparent especially at high concentrations. This might be attributed to the flexibility of the system, as the contour length $l=880$~nm of the virus is not much smaller than its persistence length $l_p=2.2$~m \cite{FrDa1970}. It can be shown, however, that the structural response to an applied flow field is chiefly governed by rod alignment \cite{ScNe2005} if the rods are sufficiently long. More importantly, the available theoretical approaches always scale the shear rate with a fixed effective rotational diffusion coefficient \cite{DhBr2003,DoEd1986,Doi1981,KuDo1983}.
The increased ordering with shear flow implies, on the other hand, that more space for the free diffusion of individual rods is created \cite{Onsa1933}, which~couples back to the flow rate needed to achieve even higher ordering. This might lead to discrepancies between theory and experiment. 
Another feature of the sheared isotropic phase is that theory predicts the formation of a biaxial para-nematic phase for dilute and semi-dilute suspensions of rigid rod-like particles under shear flow \cite{AnEd+1995}, which has received little attention in experiments so far. Biaxiality~has been observed for intrinsically-biaxial particles, such as board-like particles \cite{PeBo+1999}, and semi-flexible polymers, such as F-actin, due to the shear-induced hairpins, the orientation of which depends on shear rate and concentration \cite{KiGu+2014,HuHa+2014,Hara2013}. 

In this paper, we will address the given issues by applying rheo-SANS, which is very suited to characterize absolute orientational ordering due to the high contrast \cite{LiHe1989}, to a model system of dilute to semi-dilute fd-virus suspensions. The scattering experiments are performed in two different directions such that information in the flow-gradient (1-2) as well as the flow-vorticity (1-3) direction can be obtained. The 1-2 plane measurements, thereby, require a very subtle fine-tuning of the measurement equipment \cite{LiNe+2006,LiNe+2009}. Only by using this combination of scattering geometries, we can measure the full-dimensional orientational order given by the orientational distribution function (ODF), instead of the usual projected quantities. Whilst it is not possible to obtain the mechanical response in the 1-2 geometry, this is the case for the 1-3 geometry, such that we can couple the structural with the mechanical response. We compare the results with theoretical calculations based on Smoluchowski, as well as Doi–Edwards–Kuzuu (DEK) theory \cite{DhBr2003}. The full picture makes a calculation of novel information on system characteristics, such as biaxiality, feasible. Qualitative similarities have also been found between simulations and experiments on the isotropic-nematic transition in shear flow. This behavior is strongly influenced by the markedly different response of both phases, as the nematic undergoes a tumbling motion at low shear rates, while the isotropic phase shows flow alignment and shear thinning~\cite{RiHo+2008}.

The paper is organized as follows. First, we revisit relevant theories, describing the connection between orientational ordering, flow and mechanical response; next, we introduce the experimental methodology, including the explanation of how we obtain all relevant ordering parameters from the experiments in both scattering geometries. We then present the results and discuss the agreements and discrepancies that we find with theory.

\section{Theory}
\label{2}
\vspace{-6pt}
\subsection{Smoluchowski Theory}
\label{2.1}

The Smoluchowski theory is very suitable for the description of nonequilibrium phase behavior as it incorporates in an elegant way the external field, as well as interaction potentials, calculating the relevant equation of motion for a probability density function. In the case of colloidal rods, we are interested in their ODF $\Psi(u,t)$, which is a function of the orientation of a rod $u$ and time $t$. The N-particle Smoluchowski equation leads to \cite{DhBr2003}: 

\begin{equation}
\label{Smoluchowski}
\frac{\partial\Psi(u,t)}{\partial t}=D_r^0\mathcal{R}\cdot\left(\mathcal{R}\Psi(u,t)-\beta\Psi(u,t)\overline{T}(u,t)\right)-\dot{\gamma}\mathcal{R}\cdot \Psi(u,t)u\times(\Gamma\cdot u)~~
\end{equation}
where $D_{r}^0=3k_BT\ln(L/d)/\pi\eta_sL^3$ is the rotational diffusion coefficient in an infinitely-dilute suspension depending on the solvent shear viscosity $\eta_s$, as well as the length $L$ and thickness $d$ of the rod, and $\dot\gamma$ is the shear rate. The rotational operator is defined as $\mathcal{R}=u\times\partial/\partial u$, $\beta=1/k_BT$; the shear-field induced torque on the rods is:

\begin{equation}
\overline{T}(u,t)=-\rho\int dR \oint du'\Psi(u,t)g(R,u,u',t)\mathcal{R}U(R,u,u',t)~~
\end{equation}
where $g(R, u, u', t)$ is the pair correlation function of the rod under inspection and an additional rod with orientation $u'$ with the center-to-center distance $R=r-r'$, $U$ is the pair interaction potential of the two rods and $\rho=N/V$ is the number density. With the tensor:

\begin{equation}
\Gamma=\left[\begin{array}{ccc} 0 & 1 & 0 \\ 0 & 0 & 0 \\ 0 & 0 & 0 \end{array}\right]~~
\end{equation}
and the shear rate, we can write down the equation of motion for the orientational ordering tensor $S$ from the Smoluchowski equation:

\begin{equation}
\label{EOM}
\frac{dS}{dt}=-6D_{r}^0\left(S-\frac{1}{3}I+\frac{L}{d}\varphi(\langle uuuu\rangle:S-S\cdot S)\right)+
\end{equation}
$$~~~~+\dot\gamma(\Gamma\cdot S+S\cdot\Gamma^T-2\langle uuuu\rangle:E)$$

The orientational ordering tensor depends on the ODF in the following manner:

\begin{equation}
\label{Sdef}
S=\oint du\Psi uu=\langle uu\rangle~~
\end{equation}

In Equation~\ref{EOM}, the volume fraction of rods in the suspension is given as $\varphi=(\pi/4)d^2L\rho$, and we define the following tensor in addition to the unit tensor $I$:

\begin{equation}
E=\frac{1}{2}(\Gamma+\Gamma^T)~~
\end{equation}

To solve Equation~\ref{EOM}, we use the following closure relation \cite{DhBr2003} for both of the double contracted forms $\langle uuuu\rangle:A$ of the fourth order orientational tensor:

\begin{equation}
\label{closure}
\langle uuuu\rangle:A=\frac{1}{5}(S\cdot \overline{A}+ \overline{A}\cdot S- S\cdot S\cdot \overline{A}- \overline{A}\cdot S\cdot S+2S\cdot \overline{A}\cdot S+3SS: \overline{A})~~
\end{equation}
where $A$ stands for either $S$ or $E$ and $\overline{A}=(A+A^T)/2$. It is important to notice that in deriving the equation of motion for $S$ (Equation~\ref{EOM}) from the N-particle Smoluchowski equation for the ODF (Equation~\ref{Smoluchowski}), the shear rate-independent expression for the pair correlation function:

\begin{equation}
\label{PCF}
g(R, u, u', t)=\exp\{-\beta U(R,u,u',t)\}
\end{equation}
is used, following Onsager \cite{Onsa1933} and Doi and Edwards \cite{DoEd1986}. Here, $r$ and $r'$ denote the positions of two neighboring rods, while $u$ and $u'$ denote the orientations of their long axis. 
Assuming a hard-core interaction, this gives the excluded volume:

\begin{equation}
\label{EV}
V(u,t)=2dL^2\beta^{-1}\rho\oint du'\Psi(u',t)|u\times u'|~~
\end{equation}

The fact that there is no shear rate dependence in $g(R, u, u', t)$ is one of the main assumptions of the theory, also directly leading to the given form of Equation~\ref{EV}.

The connection of this microscopic description, using the orientational ordering tensor, to the mechanical macroscopic measurables can be made \cite{DhBr2003} via the deviatoric stress tensor $\Sigma$, given as:

\begin{equation}
\label{consti}
\Sigma=2\eta_s\dot\gamma\left(E+\frac{(L/d)^2}{3ln(L/d)}\varphi\left[\Gamma\cdot S+S\cdot \Gamma-\langle uuuu\rangle:E-\frac{1}{3}IS:E-\frac{1}{\dot\gamma}\frac{dS}{dt}\right]\right)~~
\end{equation}

The shear viscosity can be expressed in terms of the relevant component of the stress tensor using the following constitutive relation, which can be derived by plugging an expansion of the ordering tensor to the third power in shear rate into Equation~\ref{consti}:

\begin{equation}
\Sigma=2\eta\dot\gamma E+\eta_s\frac{1}{20}C\frac{\dot\gamma^2}{D_r^{coll}}\varphi\left[\begin{array}{ccc} 19 & 0 & 0 \\ 0 & -11 & 0 \\ 0 & 0 & -8 \end{array}\right]~~
\end{equation}

where $C=8(L/d)^2/45\ln(L/d)$ and the concentration-dependent collective rotational diffusion coefficient $D_r^{coll}$ is used because the ODF also depends on the distance to the isotropic-nematic transition. This links the shear viscosity with the order parameter tensor and its corresponding~eigenvalues.

Finally, also the zero-shear viscosity $\eta_0$ as a function of volume fraction $\varphi$ of rods can be calculated from this expression, leading to what is called a mixing rule for this quantity:

\begin{equation}
\label{MR1}
\eta_0=\eta_s(1+C\varphi)~~
\end{equation}

Equation~\ref{MR1} can be compared, e.g., to the theory by Berry and Russel \cite{BeRu1987}, where the polynomial for the zero-shear viscosity is of second order in $\varphi$:

\begin{equation}
\eta_0=\eta_s(1+C\varphi+\frac{2}{5}C^2\varphi^2)~~
\end{equation}

This mixing rule clearly shows a much stronger concentration dependence of the zero-shear viscosity, as was already pointed out by Dhont and Briels \cite{DhBr2003}.

\subsection{Doi–Edwards–Kuzuu (DEK) Theory}
\label{2.2}

The DEK theory constitutes a basically similar, but nonetheless very differently interpreted alternative to the theory outlined above \cite{DhBr2003}.

On the one hand, the rotational diffusion coefficient of the dilute system $D_r^0$ used in Smoluchowski theory is replaced by a purely phenomenological function $\overline{D}_r$. What is taken into account in this function can be made quantitative by the tube model \cite{DoEd1986}. Thereby, time-dependent tube disengagement of the rods leads to changed rotational relaxation times if the system is in the semi-dilute state. A~comparison of the mean square displacement of rods in the dilute state to that in the semi-dilute state directly gives a scaling law for the effective rotational diffusion coefficient. This can be written as:

\begin{equation}
\label{tube}
\overline{D_r}=\overline{C} D_r^0(\rho L^3)^{-2}~~
\end{equation}
where $\overline{C}$ is a coefficient with magnitude $10^3$ to $10^4$ \cite{DoEd1986}. According to Doi and Edwards \cite{DoEd1986}, one has to take into account this apparent concentration dependence of the rotational diffusion coefficient in the Smoluchowski Equation~\ref{Smoluchowski}. However, this description is only valid if the surrounding rods under consideration constitute a tube without possessing a preferred direction themselves. In the case of an anisotropic distribution of the surrounding, an effective mean change of the tube has to be considered if the picture of the tube as used by Doi and Edwards is believed to be a valid means of description for the given problem. Then, the function for the rotational diffusion coefficient becomes orientation dependent \cite{DoEd1986}:

\begin{equation}
\label{tubeDilation}
\overline{D}_r(u)=D_r^0\left[\frac{4}{\pi}\int du'|u\times u'|\Psi(u')\right]^{-2}~~
\end{equation}
 
Furthermore, if the external perturbation is large, one would have to take tube dilation into account. A description of such a situation is only feasible up to a certain degree of accuracy \cite{DoEd1986}. Although the DEK approach provides one of the possible non-analytic expressions for such a case, we~do not include the corresponding calculation in our treatment, since we want to follow the procedure given by Doi and Edwards \cite{DoEd1986}.

On the other hand, this description is accompanied by neglecting the pair-interaction potential $U$ in Equation~\ref{Smoluchowski} entirely. In doing so, the Smoluchowski equation with the alternative description of rotational diffusion, but in absence of interaction, is solvable analytically. Thus, one can give a direct equation for the orientational order-dependent shear viscosity:

\begin{equation}
\label{DEKequation}
\frac{\eta}{\eta_0}=\frac{\rho}{6\beta\overline{D}_r}\frac{(1-\langle P_2\rangle)^2(1+2\langle P_2\rangle)(1+3\langle P_2\rangle/2)}{(1+\langle P_2\rangle/2)^2}~~
\end{equation}
where the rotational diffusion coefficient is given by Equation~\ref{tube}, the orientational order is represented by the order parameter $\langle P_2\rangle$, which depends on the highest eigenvalue of $S$, called $\lambda_1$, as $\langle P_2\rangle=\frac{1}{2}(3\lambda_1-1)$, and $\eta_0$ is the zero-shear viscosity. It needs to be mentioned here that Equation~\ref{DEKequation} has been derived by Doi and Edwards \cite{DoEd1986} for the case of concentrated solutions. It can, however, be used for our purpose, as will be shown in section~\ref{3}. %figures should be cited in numerical order, please check and revise.

Using the measurements at hand, we are able to review both theoretical approaches from a new~perspective.
\subsection{Scaling}
\label{2.3.1}

In comparing experiments with theory, one needs to choose the right scaling for shear rate, as well as viscosity. As we indicated in the previous section, the viscosity is enslaved by the ordering in the system. In earlier experiments on worm-like systems \cite{FoKo2005}, indeed, a scaling law for the
viscosity was obtained when plotting the viscosity, scaled by the zero-shear viscosity, against the orientational order parameter $\left\langle P_2 \right\rangle $, the highest eigenvalue of a projection of $S$ into the measurement plane.

The appropriate scaling of the shear rate can be done by relating it to the rotational diffusion of the rods, $D_{r}$, such that:

\begin{equation}
\label{Peclet}
Pe=\frac{\dot{\gamma}}{D_{r}}
\end{equation}
where $Pe$ is the so-called Peclet number. The question is if one should consider the collective rotational diffusion $D_{r}^{coll}$ or the self-diffusion $D_{r}^{self}$, where we have to distinguish between the self-diffusion at infinite dilution $D_{r}^{0}$ and the self-diffusion at the volume fraction of interest, as given in Equation \ref{tube}. The right choice for this scaling will enable us to identify the interdependence of macroscopic flow behavior and microscopic orientational order, which is the focus of our experimental investigation. It should be mentioned that since the relaxation time of the rods can be given as $\tau_r=1/2D_r$ \cite{KiLo1996}, we~will expect a value of $\tau_r\dot\gamma=1$ at $Pe=2$.

The collective rotational diffusion coefficient $D_r^{coll}$ goes to zero when approaching the isotropic-nematic spinodal from a dilute, isotropic dispersion by increasing the scaled volume fraction $\frac{L}{d}\varphi$ \cite{DhBr2003}. This number should be compared to its theoretical value at the spinodal point, which can be derived by studying how a perturbation from an isotropic state $\delta S$ evolves over time, using~Equation~\ref{EOM} with the indicated closure relation and setting the shear rate to zero. This procedure leads to:

\begin{equation}
\label{orderEq}
\frac{d}{dt}\delta S=-6D_r^{coll}\delta S~~
\end{equation}
where $D_r^{coll}=D_r^0(1-L\varphi/5d)$. Therefore, the isotropic-nematic spinodal point is located at a scaled volume fraction of $\frac{L}{d}\varphi_{IN}=5$. It should be mentioned that, depending on which assumptions are made on the interaction potential $U$ between the rods, the values for this number range between $\frac{L}{d}\varphi_{IN}=4$ and $\frac{L}{d}\varphi_{IN}=5$. Furthermore, if the effective rod thickness $d_{eff}$ is greater than the actual thickness $d$, due to the surface charge, the value has to be adapted accordingly. We will, therefore, write down the relevant equations using the effective thickness.

It is to be anticipated that the scaling with $D_r^{coll}$ will not suffice, as the self-diffusion is strongly reduced at elevated concentrations. This has been nicely shown in event-driven Brownian dynamics simulations by Tao et al. \cite{TaDe2006}. From these simulations an alternative expression for $D_r^{coll}$ was found for rods of sufficient length:

\begin{equation}\label{eq_scaleShear}
\frac{D_{r}^{coll}}{D_{r}^{0}}=A(\frac{L}{d_{eff}}\varphi_{IN}-\frac{L}{d_{eff}}\varphi)^\nu~~
\end{equation}
where $\nu=(1/\sqrt{2})(L/d_{eff})^{1/4}$ and $A$ is a constant, which normalizes the function at $\varphi=0$ \cite{TaDe2006}. This~scaling takes the decrease in the rotational self-diffusion into account to some extent. We will use it for the definition of an effective Peclet number $Pe_{eff}=\dot\gamma/D_r^{coll}$ by fitting the correct value of $\varphi_{IN}$ for our system. This fit will be done such that a master curve of the order parameter, as well as scaled viscosity versus $Pe_{eff}$ is obtained, assuming that the viscosity is fully enslaved by the ordering and the ordering is fully enslaved by the collective rotational diffusion. In our theoretical predictions, we will use a corrected Peclet number $Pe_{corr}=\dot\gamma/D_r^{coll}$, where the theoretical value for $\varphi_{IN}$ will be used for defining a suitable $D_r^{coll}$.

\section{Experiments, Materials and Methods}
\label{x}
\vspace{-6pt}
\subsection{Measurements and Materials}
\label{2.4}

\begin{figure}[tbp]
\centering
\includegraphics[width=\linewidth]{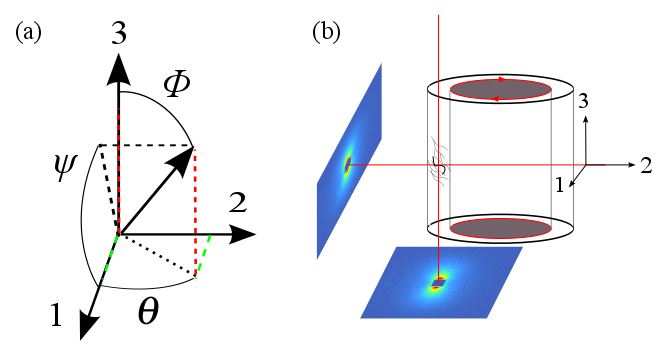}
\caption{
(\textbf{a}) Definition of angles in our scattering geometry as defined by the flow field, dashed lines are guides to the eye; (\textbf{b}) scattering planes visualized on behalf of the Couette cell, red circles indicate moving parts, red arrows show the direction of motion.}
\label{fig0}
\end{figure}

The 1-3 plane rheo-SANS and stress data were acquired simultaneously by mounting an Anton Paar MCR 501 rheometer (Anton Paar GmbH, Graz, Austria), used in the strain-controlled mode with a quartz Couette geometry, in the SANS-1 neutron beam at the SINQ spallation source at the Paul Scherrer Institute (PSI) in Villigen, Switzerland. 1-2 plane scattering experiments were carried out by using a shear cell mounted to the d22 large dynamic range small-angle diffractometer at the Institut Laue-Langevin (ILL) in Grenoble, France~\cite{LiNe+2006}. The scattering geometry is shown in Figure~\ref{fig0}b. 
All measurements were performed at~28.5~$^\circ$C.

\begin{figure}[tbp]
\centering
\includegraphics[width=0.8\linewidth]{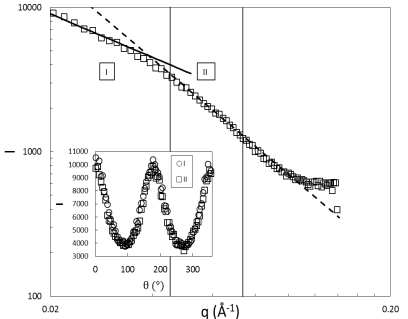}
\caption{
Porod plot for $c=11$~mg/mL without shear flow. The solid line is a $q^{-1}$ fit and the dashed line a $q^{-2}$ fit to the data. The $q$-ranges I and II of analysis are parted by the vertical lines. Inset: comparison~of the azimuthal intensity profiles for $q$-ranges I and II.}
\label{fig1}
\end{figure}

In order to find a compromise between intensity and the probed length scale of the rod, we~scanned the scalar order parameter over a range of $q$-values under stationary flow conditions using a detector distance of 6~m and wavelength of the thermal neutrons of $1.3\pm0.1$~nm. The Porod plot for a rod concentration of 11~mg/mL in absence of shear is shown in Figure~\ref{fig1}, where the rigid-rod slope $I(q)\sim q^{-1}$ is seen at quite low $q$-values, and it crosses over to a Gaussian chain slope $I(q)\sim q^{-2}$ at higher $q$. We choose the second range for our analysis and evaluate a $q$-range between 0.032 and 0.046~\AA~ as the intensity is higher in this range. However, an evaluation at smaller scattering vectors does not influence the outcome in the azimuthal averaging profile necessary for the computation of the projected order parameter $\langle P_2\rangle$, as can be seen from the overlapping of curves in the inset of Figure~\ref{fig1}. The measurement data were reduced using the GRASP freeware provided by the ILL (Grenoble, France) and subsequent analysis in Wolfram Mathematica (Wolfram, Oxford, UK) . 

While the Anton Paar MCR 501 is inherently a stress-controlled rheometer operating in strain-controlled mode, the rheological measurements presented here were replicated within experimental uncertainties using a strain-controlled ARES G2 rheometer (TA Instruments, New Castle, PA, USA).

The samples for the two different measurements were prepared (see \cite{SaRu2001}) by using a slightly different Tris buffer. The corresponding concentrations are 90 and 110~mM Tris with suspended filamentous bacteriophage fd in the following concentrations: $c\in\{4.1,11,12.43,15, 20\}$~mg/mL. Respective effective diameters of fd-virus can be considered as $d_{eff}$(90~mM)~$\approx8.64$~nm and $d_{eff}$(110~mM)~$\approx12$~nm \cite{DoFr2006}.

\subsection{Obtaining the Full Orientation Tensor and the Biaxiality}
\label{3.2}

Intensity profiles, as shown in the inset of Figure~\ref{fig1}, are directly proportional to the normalized ODFs $f(\theta)$ and $f(\psi)$ \cite{BeRo+1998}, which are projections of the overall ODF $f(\alpha)$ into the 1-2 or 1-3 plane, respectively. Thus, the director, which describes the averaged orientation $\alpha$ of the rods with the flow, makes an angle $\psi$ with the flow direction in the 1-3 plane and an angle $\theta$ with the flow direction in the 1-2 plane, as can be seen in Figure~\ref{fig0}a.

In order to parametrize the orientational ordering by single scalars \cite{BiBa+2010}, we fit the intensity profiles with a Maier--Saupe type of ODF: 

\begin{equation}\label{Eq_picken}
f(\alpha)=I_0\exp[\tilde{C} P_2(\alpha-\alpha_{max})-1]
\end{equation}

Here, $I_0$ is the amplitude, $\tilde{C}$ describes the width of the intensity profile and $P_2(\alpha-\alpha_{max})=\frac{1}{2}(3\cos(\alpha-\alpha_{max})^2-1)$ is the second order Legendre polynomial, allowing for~the director to point along $\alpha_{max}$. It has been shown earlier that this function fits the orientational distribution well, from low to high ordering \cite{HoUg+1996,PuDo+2003,Lonetti2011}.
This function can be further reduced to a single scalar that quantifies ordering, given by the average orientational order parameter $\langle P_{2}(\alpha)\rangle$, which can be calculated using the fitted parameters:

\begin{equation}
	\langle P_{2}(\alpha)\rangle=\frac{\int_0^{1}d\cos(\alpha)f(\alpha)P_2(\alpha-\alpha_{max})}
{\int_0^{1}d\cos(\alpha)f(\alpha)}
\end{equation}

In this way, the orientational ordering in this plane is parametrized by $\langle P_{2}(\alpha)\rangle$ and $\alpha_{max}$. The~procedure is identical for both parametrizations $\theta$ and $\psi$.

To relate the order parameters to the orientational ordering tensor $S$, we first note that the apparent largest eigenvalues $\lambda_1(\alpha)$ of the two plain tensor projections are related to $\langle P_{2}(\alpha)\rangle$ as $\langle P_{2}(\alpha)\rangle=\frac{1}{2}(3\lambda_1(\alpha)-1)$. In addition to the aforementioned orientation tensor $S$, we need the traceless orientation tensor $Q=(3S-I)/2$ in order to describe the relative rod positions in the reference frame of the measurement. The two tensorial quantities are related as follows:

\begin{equation}
\label{tensor}
S=(R^{(0,\theta)})^{T}\tilde{Q}R^{(0,\theta)}~~
\end{equation}
where:

\begin{equation}
R^{(0,\theta)}=\left(\begin{array}{ccc} \cos(\theta) & \sin(\theta) & 0 \\ -\sin(\theta) & \cos(\theta) & 0 \\ 0 & 0 & 1 \end{array}\right)
\end{equation}
is the rotation tensor of the laboratory reference frame and $\tilde{Q}=\frac{1}{3}(2Q+I)$ is the full orientational ordering tensor in this reference frame. We know from the measurement in the 1-3 direction that the maximum angle $\psi_{max}\equiv0$, irrespective of the applied shear rate. Therefore, the rotational tensor only depends on the angle $\theta$. From this observation and the definition of the orientation tensor with the expected symmetry conditions, one finds the following equations for the components of $Q$:

$$Q_{33}=\lambda_{1}(\theta)~~$$

\begin{equation}
Q_{22}=T-\frac{1}{2}\lambda_{1}(\theta)~~
\end{equation}

$$Q_{11}=-T-\frac{1}{2}\lambda_{1}(\theta)~~$$
where $\lambda_i(\alpha)$ is defined as the $i$-th eigenvalue of $S$ measured in the plane parametrized by the angle $\alpha$, and the biaxiality parameter $T$ can be found as:

\begin{equation}
\label{biax}
T=\frac{1}{2\left(2-\lambda_1(\theta)-\lambda_1(\psi)\right)}\left[\lambda_1(\theta)\lambda_1(\psi)-\left(\lambda_1(\theta)\right)^2\right]~~
\end{equation}

Since all of the eigenvalues $\lambda_i(\alpha)$ are measured, the traceless tensor $Q$ can be calculated from knowledge of the angle $\theta$, which defines the rotation tensor. This, of course, also gives the full orientational ordering tensors $\tilde{Q}$ and $S$. The measured order parameter tensor is evaluated at every shear rate, which corresponds to the effective Peclet number of the system in that instant.

\section{Results}
\label{3}

\begin{figure}[tbp]
\centering
\includegraphics[width=\linewidth]{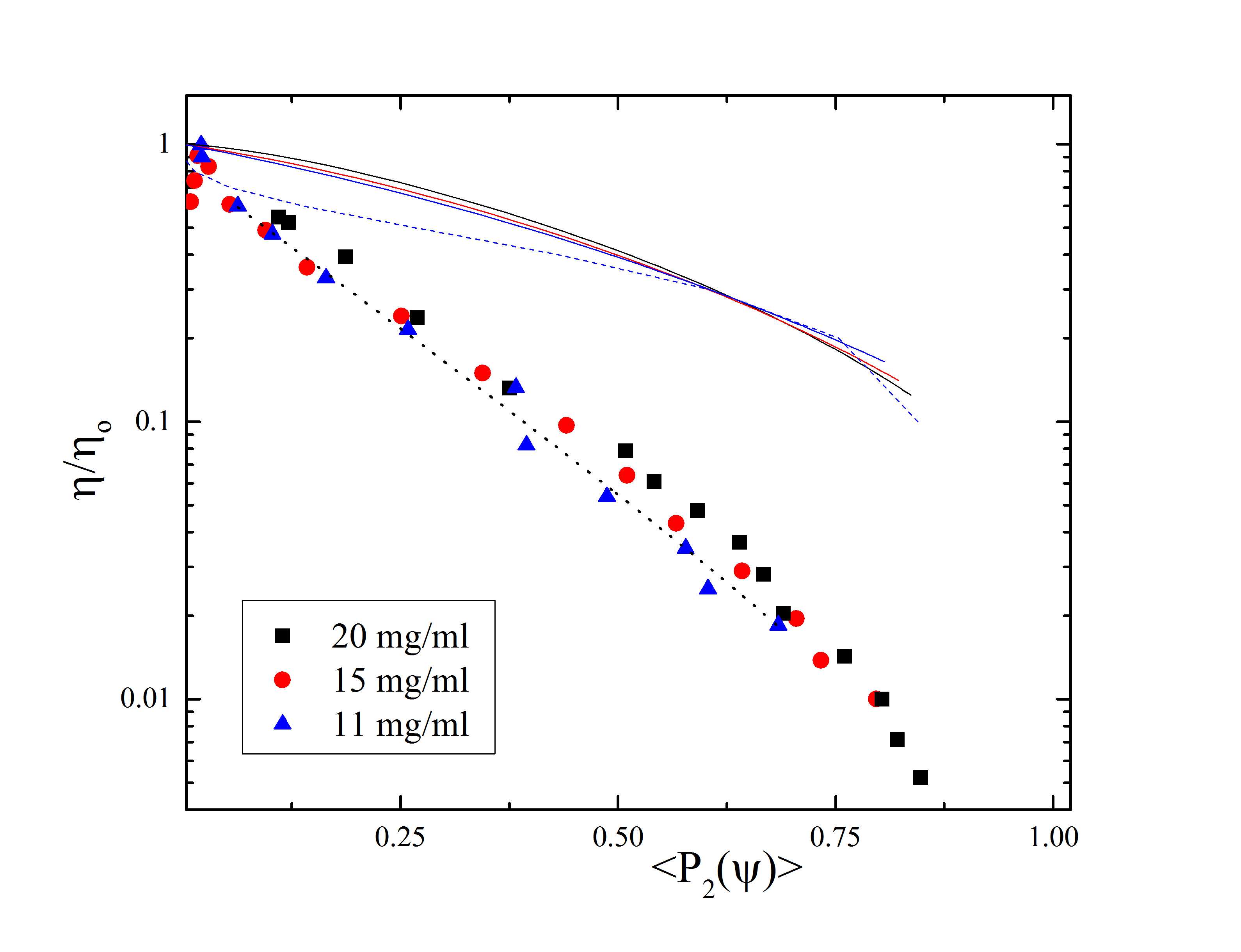}
\caption{
Corrected viscosity versus the orientational order parameter $\langle P_2(\psi)\rangle$. The full lines are calculated by Smoluchowski theory for the indicated concentrations, while the dashed line is based on Doi–Edwards–Kuzuu (DEK) theory for the smallest given concentration, and the dotted thick line is an empirical fit.}
\label{fig4}
\end{figure}

We first study the interdependence between the shear viscosity and the projected order parameter in the 1-3 plane of flow for three different rod concentrations in the dilute to semi-dilute regime. Since~the single-particle contribution to the viscosity is negligibly small, we assume that there exists a master curve for the overall comparison between shear thinning and orientational ordering, independent of the rod concentration. For obtaining the master plot, one would ideally scale the viscosity by the zero-shear viscosity $\eta_0$, but for this system of long rods and for this rheometer, it is not feasible to obtain the zero-shear viscosity, as shear thinning sets in immediately. Therefore, we~scaled the viscosity with an adjustable $\eta_0$, such that the data collapse on a single curve in a log-lin plot; see~Figure~\ref{fig4}. For each concentration, we thus find a zero-shear viscosity as the correction value of the shear viscosity in Figure~\ref{fig4}.

\begin{figure}[tbp]
\centering
\includegraphics[width=\linewidth]{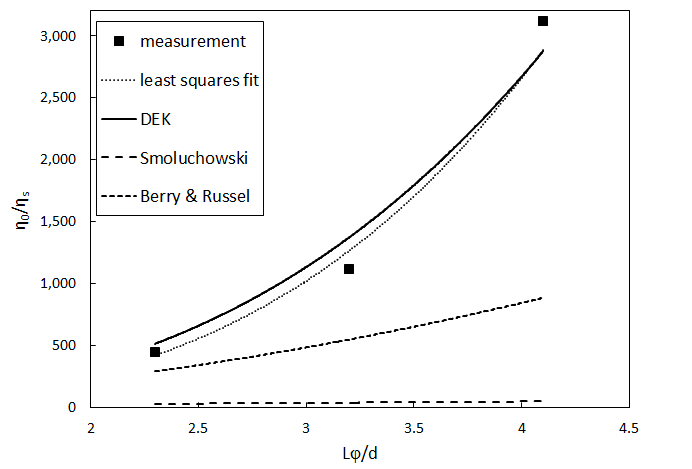}
\caption{Zero-shear viscosity scaled by the solvent viscosity as a function of the volume fraction from data shifting. The dotted line is a least squares log-linear fit to the data points, while the other lines represent the different theories.}
\label{fig3}
\end{figure}

In Figure~\ref{fig3} we compare the thus obtained $\eta_0$, scaled by the solvent shear viscosity, with Smoluchowski, as well as DEK theory and a comparable theory by Berry and Russel \cite{BeRu1987}. It is seen that neither the Smoluchowski theory nor the theory by Berry and Russel fit the data. However, DEK~theory approximates the data reasonably well.

In Figure~\ref{fig4}, we also plot the theoretical predictions as full and dashed lines. The discrepancy between theory and experiment in this representation is evident, and neither DEK nor Smoluchowski theory values predict the actual experimental outcome. Fitting the order parameter-dependent corrected viscosity curves with a purely empirical exponential form, suggested by F\"orster et al. \cite{FoKo2005} (see the dotted thick line in Figure~\ref{fig4}), we find that: 

\begin{equation}\label{eq_scaleEta}
\frac{\eta}{\eta_0}=\exp\left\{-a\langle P_2(\psi)\rangle\right\}
\end{equation}
describes the data well when $a=5.0$.\par

\begin{figure}[tbp]
\centering
\includegraphics[width=\linewidth]{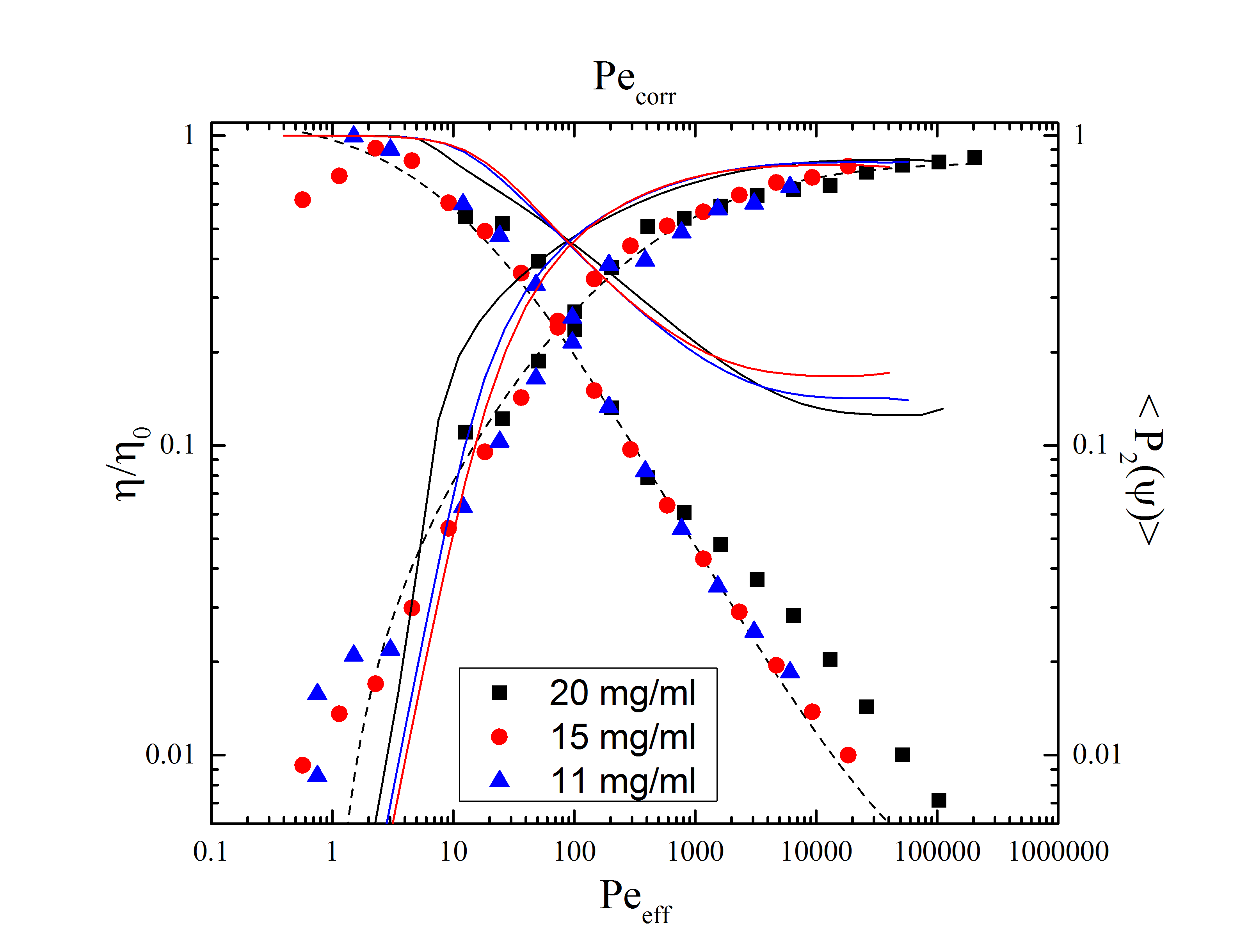}
\caption{Shear alignment given by the orientational order parameter $\langle P_2\rangle$, as well as shear thinning given by $\eta_{corr}:=\eta/\eta_0$ versus the effective Peclet number for the given concentrations. Theoretical curves are given as solid lines, and the dashed line is a least squares fit of the measurement.}
\label{fig2}
\end{figure}

Next, we focus on the relation between shear viscosity and shear rate. A master plot can be obtained for $\eta_{corr}=\eta/\eta_0$ versus the shear rate, when defining an effective Peclet number $Pe_{eff}$. The~scaling of the shear rate works well when using $Pe_{eff}=\dot\gamma/D_r^{coll}$, where $D_r^{coll}$ is given by Equation~\ref{eq_scaleShear} and choosing $\frac{L}{d}\varphi_{IN}=4.2$; see Figure~\ref{fig2}. The focus of this scaling lies on the higher shear rate regime, due to the low torques at low shear rates. The same scaling of the shear rate is used when plotting the projected order parameter $\langle P_2(\psi)\rangle$ (in the 1-3 plane) as a function of the imposed shear rate. Again, a collapse of data is achieved by the same Peclet numbers as above; see Figure~\ref{fig2}. Note~that for the theoretical data to fall on top of each other, we need to use the theoretical $\frac{L}{d}\varphi_{IN}=5$, giving~a corrected Peclet number $Pe_{corr}$ (see Section~\ref{2.3.1}), while the scaling law by Tao \cite{TaDe2006}, Equation~\ref{eq_scaleShear}, remains valid in its given form.

\begin{figure}[tbp]
\centering
\includegraphics[width=\linewidth]{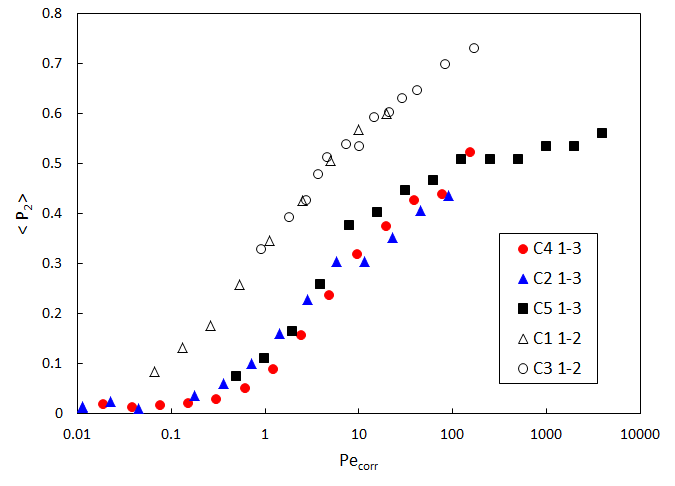}
\caption{Projected order parameter as a function of the (effective thickness-) corrected Peclet number for the measured concentrations; the numbers $j$--$k$ indicate the unit vectors in the measurement plane.}
\label{fig5}
\end{figure}

As the master curve with the right scaling of the shear rate works very well for the orientational ordering, as well as for the viscosity, using $\frac{L}{d}\varphi_{IN}=4.2$, we now proceed to compare the behavior of rods in the two measurement directions. We apply the same Peclet number scaling to the projected order parameter in the 1-2 plane, as well as in the 1-3 plane; see Figure~\ref{fig5}. Before doing so, we need to correct for the disparate effective thicknesses in the two measurements, see Sections~\ref{2.3.1} and \ref{2.4}, as well as Equation~\ref{eq_scaleShear}.

Again, a data collapse for the curves of different rod concentrations in both measurement directions can be achieved. It is seen that the observed orientational order parameter in the 1-2 plane is higher than the one in the 1-3 plane from the very beginning of shear-induced order, corresponding to the low Peclet number regime. The discrepancy grows non-linearly with $Pe_{eff}$ and shows the largest differences in the high Peclet number regime.

\begin{figure}[tbp]
\centering
\includegraphics[width=\linewidth]{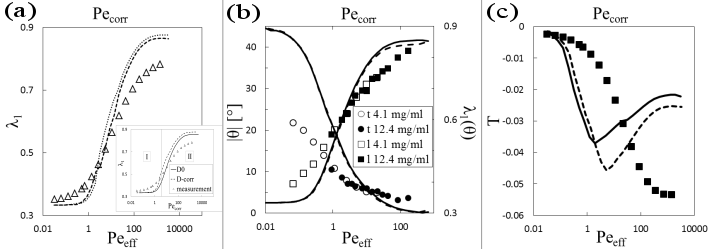}
\caption{(\textbf{a}) Measured largest eigenvalue of $S$ as a function of the corrected Peclet number compared to the Smoluchowski theory, evaluated for two effective thickness values $d_{eff}^{(1-3)}=8.6$ (dotted line) and $d_{eff}^{(1-2)}=12$ (dashed line) at a concentration of 11~mg/mL. Inset: largest eigenvalue compared to Smoluchowski theory for two different rotational diffusion coefficients. Two regimes are marked, separated at $\tau_r\dot\gamma=1$.
(\textbf{b}) Measured azimuthal tilt angle $\theta$ [t] and largest projected order tensor eigenvalue $\lambda_1(\theta)$ [l] as a function of the corrected Peclet number, for two concentrations, compared to the Smoluchowski theory. The solid and dashed lines are theoretical curves for the two given concentrations.
(\textbf{c}) Measured biaxiality parameter $T$ (squares) versus corrected Peclet number compared to Smoluchowski theory for the two effective thicknesses $d_{eff}^{(1-3)}=8.6$ (solid line) and $d_{eff}^{(1-2)}=12$ (dashed line).}
\label{fig7}
\end{figure}

The 1-2 plane measurements provide additional information on the achievable order in the system, as they also give a value for the angle $\theta$ the director makes with the flow direction, as plotted in Figure~\ref{fig7}b. With the appropriate identified scaling of the shear rate, the orientational order parameters $\langle P_2 (\theta)\rangle$ and $\langle P_2 (\psi)\rangle$ and the orientation of the director $\theta$ (given that $\psi_{max}\equiv0$), we have the necessary information at hand to identify all of the parameters that describe the orientational ordering, namely $\theta$, the highest eigenvalue $\lambda_1$ and the biaxiality $T$, using Equations~\ref{tensor} and \ref{biax} in Section~\ref{3.2}. These~results can be compared to the theoretical prediction from Smoluchowski theory; see Figure~\ref{fig7}a,b. It is important to notice that the Peclet numbers of two measurements for different concentrations are not equal, due to the used scaling. Therefore, the full tensorial order can only be found by fitting the relevant data $|\theta| (Pe_{eff})$, shown in Figure~\ref{fig7}b, with an appropriate function and, thus, smoothing out the apparent discrepancies without having the need for averaging. The used fitting function is a Cole-Cole type function \cite{CoCo1941}, providing a sigmoid shape in a log-log representation.

As expected, the angle $\theta$ is indirectly proportional to the order parameter and decreases with increasing shear rate. The theoretical prediction starts from a higher mean angle of approximately $|\theta|=45^\circ$ and has a much steeper decay than the experimental finding. Theory also ultimately predicts a total decay of the angle at very high shear rates. However, measurements cannot be conducted up to values where the mean angle of the rods with the flow direction is effectively zero. Similarly,~$\lambda_1$~does not reach the high ordering as predicted in theory for high Peclet numbers, while at low Peclet numbers, $\lambda_1$ starts to increase at much lower Peclet numbers than predicted by theory; see Figure \ref{fig7}a.

The biaxiality parameter $T$ of the system as a function of the corrected Peclet number is plotted in Figure~\ref{fig7}c. In experiment, biaxiality is observed as soon as ordering occurs, and it grows together with $\lambda_1$. The theoretical prediction, on the other hand, displays an increasing biaxiality only up to a certain point, corresponding roughly to $d^2\lambda_1/dPe_{corr}^2=0$ and a decay in biaxiality after this point until a certain constant value is reached, which remains up to very high imposed shear rates.

\section{Discussion}
\label{4}

\vspace{-6pt}
\subsection{Shear Thinning}
As can be seen from Figure~\ref{fig2}, there exists a universal scaling for the Peclet number-dependent order parameter if the rotational diffusion that is used to scale the shear rate has the proper concentration dependence. For the shear viscosity versus shear rate data, the same scaling can be applied. In order to have a data collapse here, the shear viscosity needs to be shifted by using the zero-shear viscosity. This~procedure leads to a master curve for shear thinning, indicating two important aspects of the rheological behavior of dilute and semidilute suspensions of long, thin, mono-disperse rods:

First, we found in Figure~\ref{3} that the dependence of the zero-shear viscosity on the relative volume fraction of rods, however few points we have, can be best described by a non-linear polynomial mixing rule of the form:

\begin{equation}
\frac{\eta_0}{\eta_s}=1+\frac{\pi}{90ln(\overline{r})}\rho L^3+\frac{\pi}{30\overline{C} ln(\overline{r})}\left(\rho L^3\right)^3~~
\end{equation}
where $\overline{C}=5800$ and $\overline{r}=L/d$ is the aspect-ratio of the viruses. This mixing rule has been derived by Doi and Edwards \cite{DoEd1986} using Equation~\ref{Smoluchowski}, setting $U=0$ and inserting an empirical rotational diffusion coefficient, which depends itself on the relative volume fraction of rods: $\overline{D_r}=\overline{C} D_r^0(\rho L^3)^{-2}$, as briefly described in the theoretical Section~\ref{2.2}. The given relation stands in marked difference to mixing rules derived via Smoluchowski theory, laid out above, which yields a first order polynomial for $\eta_0(\varphi)/\eta_s$, or the theory by Berry and Russel \cite{BeRu1987}, which yields a second order polynomial. In~our experimental investigation, the concentration dependence of the apparent rotational diffusion coefficient is slightly larger than the predicted power of $-2$ dependence by Doi and Edwards \cite{DoEd1986}. This result could be compared to numerical simulations \cite{MaDa1986}, where the exponent was found to be slightly smaller than $-2$, as well as experiments \cite{GrKr+1992}, where the dependency was seen to be even less than in the mentioned~simulations.

One can, therefore, proceed with the criticism of the existing theoretical approaches given here in two different directions, always assuming that the given closure relation in Equation~\ref{closure} assures a good accuracy, as has been shown by Dhont and Briels \cite{DhBr2003}. It should, nonetheless, be mentioned here that there remains a possibility that by coming close to the concentration of the isotropic-nematic transition, the closure relation Equation~\ref{closure} becomes less and less accurate, as it has been shown by Kr\"oger~et~al.~\cite{KrAm+2008} that the used closure does not give as good results as other closures in the nematic phase. Since~the discrepancies between theory and experiment are much bigger than the differences in closure accuracy, we will not discuss the associated issues, but instead focus on the fundamental differences between the shown theoretical approaches. Either the problem is seen in the definition of a correct rotational diffusion coefficient, which can be thought of as depending on both, the shear rate and the concentration, or we attribute the huge discrepancies to the definition of a correct pair correlation function, which should be a functional of the imposed field in non-equilibrium. In order to give this reasoning a sound grounding, experiments would be needed that probe the dynamics of rods under shear in order to test the variability of the rotational diffusion coefficient under the important influence parameters. This, however, is not a trivial task.

Since the given difference between Smoluchowski theory as we use it here and the approach by Doi and Edwards is solely attributable to the energy, which is dissipated by ordering of the rods, we single out the used pair correlation function as the weakness of the given theoretical approach. Our conclusion is based on the premise that the rotational diffusion coefficient in the small shear rate regime should not depend on the shear rate, as is assumed in the theory of Doi and Edwards by purely phenomenological reasoning. Instead, given the rotational diffusion coefficient of the infinite dilute regime $D_r^0$ as a state-independent coefficient in Equation~\ref{Smoluchowski}, a dynamic relation for the pair correlation function is needed, which leads to a higher polynomial order in the mixing rule.

Second, the corrected shear viscosity $\eta_{corr}$ strongly decreases while the orientational ordering strongly increases with increasing shear rate; see Figure \ref{fig2}. Given the fact that the zero-shear viscosity is much higher than predicted from Smoluchowski theory, this is not surprising. The effect is especially pronounced at very low shear rates, where ordering, as well as the orientation of director $\theta$ immediately jumps from 1/3 to $\lambda_1\approx0.4$, while $\theta$ jumps from $45^\circ$ to $\theta\approx 25^\circ$; see Figure~\ref{fig7}. This means that shear thinning, as a rheological signature of our system of rigid mono-disperse rods, is most significantly caused by immediate strong ordering of the rods towards the flow direction.

In addition, we observe that the angle $\theta$, which the rod director confined to the 1-2 plane makes with the flow direction, displays a power-law dependence on the shear rate, as can be seen from Figure~\ref{fig7}b, which also has been observed in multi-particle collision dynamics simulations \cite{WiMu+2004}. We~find that the power-law is independent of the concentration of our suspension if we scale the Peclet number in accordance with Equation~\ref{eq_scaleShear}. The theoretical prediction of a fully-decayed angle at high shear differs from the measurement, which shows that the angle does not reach this plateau also at the highest measurable shear rates, as can be seen from Figure~\ref{fig7}b.

Again, the origin of the discrepancy between theory and experiment in Figure~\ref{fig4} should be looked for in the shear-dependent dynamic pair correlation function. Clearly, the non-equilibrium behavior we observe here cannot be accounted for by Equation~\ref{PCF} as has been already pointed out in an earlier paper \cite{DhBr2003}. Since the expression that Doi and Edwards derive for their empirical rotational diffusion coefficient in this non-equilibrium case has not been used in their outlined analytic approach \cite{DoEd1986} (see~Section~\ref{2.2}), their prediction fails along the same lines as the Smoluchowski theory.

For high shear rates, so at the moment the tube is fully dilated and there are very little interactions between the rods, the correct scaling of the shear rate should in principle not be done by using $D_r^{coll}$, but rather by using $D_r^{0}$. This is seen from evaluating the largest eigenvalue of the orientational ordering tensor as a function of the bare Peclet number, using the rotational diffusion coefficient of the infinite dilute case, shown in the inset of Figure~\ref{fig7}a. When doing so, the theoretical values in the large shear rate regime (II in the inset in Figure~\ref{fig7}a) are closer to the experiments than what is seen in Figure~\ref{fig7}a. This might also explain why in the experiments, $\lambda_1$ and $\theta$ never reach a plateau and why a quantitative match of theory with experiments of nematic rod-like colloids of the same kind in shear has been found \cite{LeDo+2005}. The reason is that dilation does not play an important role there. Furthermore,~this~tells us that there exist two distinct regimes in the viscosity versus shear rate curve: a low shear rate regime where the rotational diffusion coefficient strongly depends on the volume fraction, as~discussed before, and a shear rate regime that does not show this dependence.

\subsection{Biaxiality}
By using the rheo-SANS measurements in two scattering geometries, we are able to obtain the full ordering tensor $Q$ from which, after diagonalization and rotation, the biaxiality parameter $T$ can be calculated. In Figure~\ref{fig7}c, we see a growing biaxiality with growing order, which goes to a plateau value at a Peclet number of approximately $Pe_{eff}\approx600$, always following the behavior of the largest eigenvalue of $S$. It comes as no surprise that the dispersions of rods show biaxiality, since~order in the gradient direction is forced by the velocity gradient in shear flow, while the exchange of momentum in the vorticity direction is small. The same observation has already been made, however,with rods of a smaller aspect-ratio, in multi-particle collision dynamics simulations \cite{WiMu+2004}. Also for living polymers, which posses characteristics of rods, as well as soft polymers, biaxiality could be seen to occur, although the orientation of the ordering plane there depends on the concentration and is due to hairpin formation, which does not play a role for the very stiff fd-virus \cite{Hara2013,KiGu+2014}. That there is order in the vorticity direction at all is, thus, already equivalent to calling the system biaxial. The projection of an order parameter in the $1-k$ plane, where $k$ is either two or three, into the perpendicular plane always gives a different result than measuring the order in the plane itself, which can be seen from Figure~\ref{fig5}.

The Smoluchowski theory predicts a shear dependence of the biaxiality parameter $T$ displaying a maximum that is roughly located at the Peclet number where the rotational relaxation time of the rods is equal to the reciprocal shear rate, which is also the inflection point of the curve for $\lambda_1$ in Figure~\ref{fig7}a. The peak and location of the maximum of $T$ slightly depends on the thickness of the rods. This is not the case for the largest eigenvalue $\lambda_1$, as is seen in Figure\ref{fig7}c. 

In contrast to the experimentally-determined shear rate dependence of $\lambda_1$ and $\theta$, the experimentally-determined biaxiality is much less sensitive to shear flow at low shear rates than predicted by theory. This suggests that due to the high degree of entanglement, torque is also transmitted in the vorticity direction, leading to an overall flow alignment.
The fact that we merely see a monotonic increase of $T$ and no return to a monoaxial distribution might also be connected to the observation that both $\lambda_1$ and $\theta$ do not reach their plateau at high shear rates and as such to the dilation of the cage. Again,~at~this high degree of ordering, the shear rate should rather be scaled by the infinite dilute rotational diffusion~coefficient.

\section{Conclusions}
\label{5}

The complete shear rate-dependent ordering tensor of sheared isotropic dispersions of quasi-ideal colloidal rods was obtained by combining two rheo-SANS geometries. Thus, we obtained the orientational order parameter $\lambda_1$, the angle of the corresponding eigenvector $\theta$ and the biaxiality $T$, which could be related to the shear thinning viscosity. In order to compare results to theory, we~scaled the shear rate with an effective rotational diffusion coefficient that approaches zero at the isotropic-nematic transition. While the process of ordering is qualitatively predicted by DEK, as well as Smoluchowski theory, there are some important differences that hint to the need for theoretical~improvements. 

At very low shear rates, all theoretical predictions underestimate the zero-shear viscosity and corresponding strong shear thinning, as well as the strong shear rate dependence of the orientational order parameter $\lambda_1$ and average angle $\theta$, which exemplifies also the interdependence between ordering and shear thinning.
We attributed this to the negligence of concentration dependence in the rotational diffusion coefficient for the low shear rate regime, as well as the use of an equilibrium pair-correlation function. While the first of these issues is apparently well-covered by DEK theory, the second has not received enough attention given that the main difficulties in the high shear rate region could be shown to lie in this function.

At high shear rates, we observe that the applied shear rates were not high enough for $\lambda_1$ and $\theta$ to reach their plateau, while $T$ did not turn back to a monoaxial value of zero. This suggests that for a high degree of ordering a different scaling of the shear rate is needed, probably with the diffusion coefficient at infinite dilution, as the local surrounding of a rod has completely changed.

\vspace{6pt} 

\vspace{+6pt}
\textbf{Acknowledgments}\\
The research is funded by the European Union within the Horizon 2020 project under the DiStruc Marie Sk\l{}odowska Curie innovative training network; Grant Agreement No. 641839.

\textbf{Author Contributions}\\ 
Christian Lang wrote the paper and analyzed the data. Minne Paul Lettinga conceived of, designed and performed the experiments under technical support by  Joachim Kohlbrecher and Lionel Porcar.

\section*{\noindent Abbreviations}\noindent 
The following abbreviations are used in this manuscript:\\\\
\vspace{6pt}
\hspace{-0.35cm}
\begin{tabular}{ll}
ODF & orientational distribution function\\
DEK & Doi--Edwards--Kuzuu\\
SANS & small angle neutron scattering\\
\end{tabular}

\end{document}